\documentclass[
aps,%
12pt,%
final,%
notitlepage,%
oneside,%
onecolumn,%
nobibnotes,%
nofootinbib,%
noshowpacs,%
centertags]%
{revtex4}
\usepackage{graphicx}

\begin{document}

\title{Ultra-High Energy Cosmic Rays from the Galactic Center}

\author{\bf \firstname{V. N.}~\surname{Zirakashvili }$^{1)}$}
\email{zirak@izmiran.ru}
\author{\bf \firstname{S. I.}~\surname{Rogovaya }$^{1)}$}

\affiliation{$^{1)}$Pushkov Institute of Terrestrial Magnetism, Ionosphere
and Radiowave Propagation, Russian Academy of Sciences, Moscow, Troitsk, Russia}%

\received{July 15, 2025} \revised{July 31, 2025} \accepted{July 31, 2025}

\begin{abstract}
\noindent
{\bf Abstract}---It is shown that Eddington-like accretion event in the Galactic center several
 million years ago and particle acceleration at accompanying shocks and jets could explain the
observed cosmic ray spectrum at energies above 1 PeV. Cosmic ray particles are confined in extended
 (several hundred kiloparsec in size) Galactic halo. It is shown that the halo magnetic field could be as small as
 $2\times 10^{-7}$ G for the effective confinement.

\end{abstract}

\maketitle

\section{Introduction}
\label{sec:intro}

It is believed now that some powerful energetic event related with
the central supermassive black hole (SMBH) occurred several million
years ago in the Galactic center (GC). It is known that radiation of
this event produced ionization cones observed in the Magellanic
stream with the estimated age of $3\pm 1$ Myrs
\cite{blandhawthorn19}. In addition young massive stars of the age
of $\sim $6 Myrs are observed inside the central parsec of GC.
Probably they were formed in the dense shell expelled during the
event \cite{nayakshin18}.  Recently discovered Fermi and e-Rosita
bubbles \cite{su10,predehl20} are probably related with shocks
produced during the event.

On the other hand relativistic jets and winds driven by the gas accretion onto central black holes
in active galactic nuclei (AGN) are
considered as possible candidates for the production of ultra-high energy cosmic rays (UHECRs)
(see \cite{bykov12} for a review). If so the  particles accelerated during the energetic event
in GC could be observed
at the Earth now. Such a model of UHECR origin was already considered \cite{kulikov69}.
The corresponding intensity of particles depends on their confinement in the Galaxy and its
surroundings.

It is known now that all galaxies have large extended halos with hot gas
and magnetic fields. They
 were created during the galactic formation and consequent star-forming and central SMBH activity. These
 halos could provide confinement of UHECRs for several million years.
In the recent paper \cite{zirakashvili24} we investigated this scenario and found that the effective
confinement is possible if the halo magnetic field strength is higher than $0.5\mu $G.

High energy particles could be accelerated at the outer bow shock driven by the jet or relativistic wind outflow
via the diffusive shock acceleration (DSA) mechanism \cite{krymsky77,bell78,axford77,blandford78}
 with the energy distribution close to $E^{-2}$.

The different spectra are expected for particles accelerated in the
jet itself. Acceleration of particles in the shear flow
\cite{berezhko81, earl88} is one of the possibilities. Probably the
jets with strong toroidal magnetic fields produce the highest energy
particles while the maximum energy of particles accelerated by bow
shocks is lower.

In the present paper, we consider our model in more detail. We shall show that the magnetic
field strength could be as small as $\sim $0.2$\mu $G in the case of heavy UHECR composition and recent
 accretion event at 6 Myrs ago.

The paper is organized as follows. In the next Section 2, we briefly remind our model. Numerical results are presented in Section 3.
The discussion and conclusions are presented in Section 4.

\section{Description of the model}

\begin{table}[tbp]
\caption{Parameters of the source components in the Galactic center}
\centering
\begin{tabular}{|c|c|c|c|p{6.0cm}|}
\hline  Component&$\gamma $&$\epsilon _{\max }$  & $E_{\mathrm{cr}}(E>1\ \mathrm{GeV})$ &
$k(A)/k_{\odot }(A)$\\
\hline   jet     &  0.0    &$3\times 10^{18}$ eV       &$2.9\times 10^{53}$ erg&
$1, A=1;\ 2,  A=4;\ 2\times 20, \ A>4$\\
\hline  bow shock &  2.0    &$4\times 10^{15}$ eV&$1.9\times 10^{55}$ erg&
$1, A=1;\ 2, A=4;\ A/4, A>16$; $2A/Z,\ 4<A\le 16$\\
\hline  inner jet   &  2.0    &$3\times 10^{18}$ eV&$3.5\times 10^{54}$ erg&$1,\ A=1;\ 0, \ A>1$\\
\hline
\end{tabular}

\end{table}

A detailed description of our model can be found in our paper \cite{zirakashvili24}.
The calculations of cosmic ray
propagation include the spatial diffusion, energy losses, and nuclei fragmentation of protons
and nuclei traveling from the central instantaneous point source. The source produces three
components of accelerated particles. Each component has a spectrum that is described by the
equation

\begin{equation}
   q(\epsilon ,A)\propto \frac {k(A)}{\epsilon ^2}\left( \frac {A\epsilon }Z\right) ^{-\gamma +2}
\exp {\left( -\frac {A\epsilon }{Z\epsilon _{\max }}\right) }, 
\end{equation}
where $\epsilon $ is the energy per nucleon, $A$ and $Z$ are the atomic mass and charge numbers respectively,
the function $k(A)$ describes the source
chemical composition and can be written in terms
of the solar composition $k_{\odot }(A)$.


To describe particle diffusion, we use an analytical approximation of the diffusion coefficient
in an isotropic random magnetic field with the Kolmogorov spectrum \cite{harari14}:

\begin{equation}
D= \frac {cl_c}{3}\left( 4\frac{E^2}{E^2_c}+0.9\frac{E}{E_c}
+0.23\frac{E^{1/3}}{E^{1/3}_c}\right) , \ E_c=ZeBl_c=0.9\ \mathrm{EeV}\ ZB_{\mu \mathrm{G}}l_{c,\mathrm{kpc}},
\end{equation}
where $E$ is the energy of the particle, $B$ is the magnetic field strength and $l_c$
is the correlation length of the magnetic field. At large energies $E\gg E_c$ the scattering of particles occurs on
 the magnetic inhomogeneities with scales smaller than the particle gyroradius and
the diffusion coefficient is proportional to $E^2$. At lower energies $E\ll E_c$ the resonant scattering results in
 the energy dependence of diffusion $\sim $$E^{1/3}$.

\section{Numerical results}

We model the propagation of particles in the spherical simulation domain with radius
 $R=400$ kpc where an absorbing boundary condition is set. The age $T$ of the Fermi and e-Rosita
bubbles considered as a result of the last active phase is not exactly known. We analyze the
model of bubble formation at $T = 6$ million years ago. The
parameters of the source spectrum are adjusted to reproduce observations and are given in
Table 1.

\begin{figure}
\begin{center}
\includegraphics[width=7.0cm]{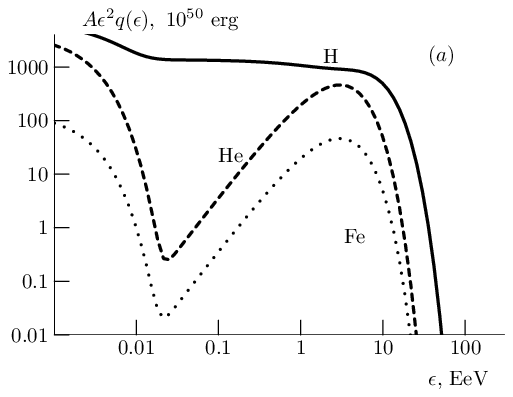}
\hfill
\includegraphics[width=7.0cm]{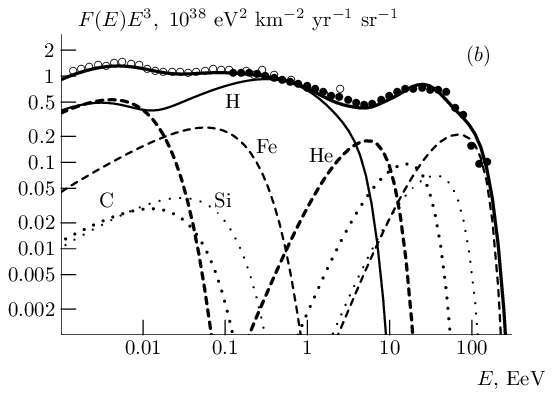}
\end{center}
\caption{Source spectra of protons (solid line), He nuclei (dashed line) and Iron (dotted line) produced
in  the Galactic center $(a)$. Spectra of different elements and all-particle spectrum
(thick solid line) produced in  Galactic center and observed at the
Earth position $(b)$. Spectra of Tunka-25, Tunka-133 array (\cite{budnev20}, open circles) and PAO
(\cite{PAO21}, energy shift +10$\% $, black circles) are also
shown $(b)$.}
\end{figure}

\begin{figure}
\begin{center}
\includegraphics[width=7.0cm]{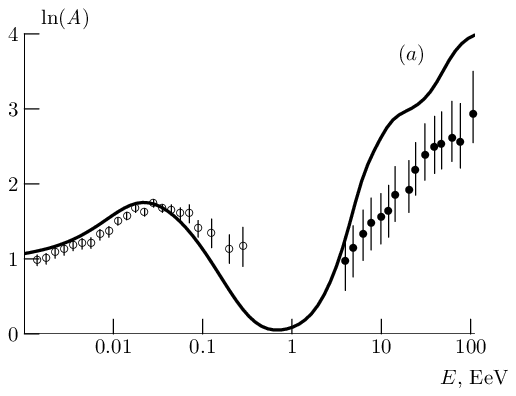}
\includegraphics[width=7.0cm]{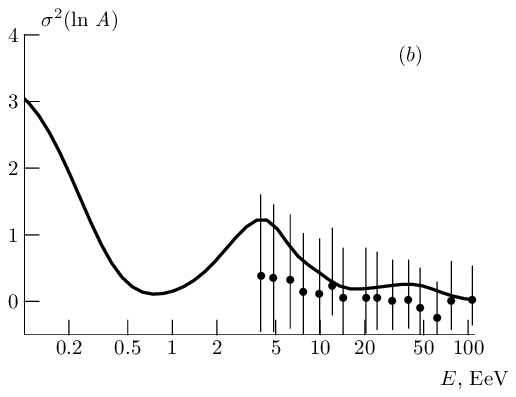}
\end{center}
\caption{Calculated mean logarithm of atomic number $A$ $(a)$ and its
variance $(b)$. The
measurements of Tunka-133, TAIGA-HiSCORE
 array (\cite{prosin22} hadronic interaction model QGSJetII-04, open circles) and PAO
(EPOS-LHC interaction model, black circles,
energy shift +10$\% $ \cite{PAO25}) are also shown. }
\end{figure}

It is expected that the bow shock and inner jet components
are produced via the diffusive
 shock acceleration mechanism while the highest energy hard jet component
 is produced via the shear acceleration mechanism.

The abundance of heavy
nuclei is adjusted to be 20 times higher in the jet in comparison with the solar composition
(see Table 1). In addition we
 assumed that the  injection is proportional to the  mass to charge ratio. This gives the additional
 factor 2 for fully ionized ions.

The third pure proton component is probably generated close to the jet origin where strong radiation
field results in the production of secondary neutrons which later turn into protons.

The total energy of the components $E_{\mathrm{cr}}$ is given in the 4th column of the Table 1.

The random magnetic field
strength $B=0.2\mu $G and the correlation length $l_c=40$ kpc are
accepted. The value of parameter $E_c$ in Eq. (2) is $E_c=7$ EeV. The corresponding free scattering path
$\lambda $ is $\lambda=200$ kpc for protons of this energy.

The source spectra of protons and nuclei are shown in left panel of Figure 1.




The spectra observed at the Solar system location at the galactocentric distance $r=R_{\odot }=8.5$ kpc
 are shown in right panel of Fig. 1. The mean logarithm and its variance are shown in Fig. 2.

The calculated degree of anisotropy
$\delta \sim 5\times 10^{-3}$ is close to $\delta =1.5R_{\odot }/cT$ which is the
anisotropy of the instantaneous point source in the infinite space.

\section{Discussion and Conclusion}

The electric potential difference is an upper limit for the maximum energy.

For protons it is given by
\begin{equation}
E^j_{\max }=e\frac uc\int ^{R_j}_0drB(r)=\frac {u}{2c}eB_jR_j .
\end{equation}
Here $u$ and $R_j$ are the jet speed and radius respectively. It was
assumed that the toroidal 
 magnetic field of the jet $B(r)$ is proportional to the distance $r$ from the jet axis $B(r)=B_jr/{R_j}$.
It is convenient to express the magnetic field strength $B_j$ via the magnetic luminosity of the
 jet $L_\mathrm{mag}$ that is the total Pointing flux of two opposite jets

\begin{equation}
L_\mathrm{mag}=u\int ^{R_j}_0rdrB^2(r)=\frac u4B^2_jR_j^2 .
\end{equation}

This gives the expression

\[
E^j_{\max }=e\sqrt{\beta _jL_{\mathrm{mag}}c^{-1}}
\]
\begin{equation}
=1.73\times {10^{19}} \ \mathrm{eV} \ \beta _j^{1/2} \left(
\frac {L_\mathrm{mag}}{10^{44}\ \mathrm{erg}\ \mathrm{s}^{-1}} \right)
^{1/2},
\end{equation}
where $\beta _j=u/c$ is the ratio of the jet
speed to the speed of light.

The magnetic luminosity of the jet is probably a small fraction (say $\sim $$10^{-2}$) of the total jet luminosity.
Therefore the jet luminosity should be of the order $10^{45}$ erg s$^{-1}$ to accelerate protons up to
several EeV in our model (see the Table 1). Note that the Eddington luminosity of Milky Way SMBH is
$5\times 10^{44}$ erg s$^{-1}$.

We see that magnetic field strength $B=0.2 \mu$G is enough for confinement of UHECRs.
Note that the mean magnetic field strength of $0.5\mu $G along the line of sight at 100 kpc
galactocentric distances was estimated from the recent measurements of the Faraday rotation
performed for different samples of galaxies \cite{heesen23,bockmann23}. So the value we used is in accordance
with observations.

The enrichment of heavy elements $\sim $20 times in the jet  is needed in our model.
This seems to be also in agreement
with spectroscopic observations of AGN. It is known that the gas metallicity derived for AGN
with Eddington accretion is higher than 10 \cite{floris24}.

We conclude that cosmic ray spectrum at energies above 1 PeV can be reproduced in our model.
At lower multi-TeV energies the transition to cosmic rays produced in other Galactic sources
(which are probably supernova remnants) could occur (see our paper \cite{zirakashvili25} for details).

\begin{acknowledgments}
The work was partly performed at the Unique scientific installation
"Astrophysical Complex of MSU-ISU" (agreement EB-075-15-2021-675).
\end{acknowledgments}

%

\end{document}